\documentclass[aps,prl,reprint,longbibliography,superscriptaddress,amsmath,amssymb,nofootinbib,nobibnotes,floatfix]{revtex4-2}
 
\usepackage{bm}
\usepackage{graphicx}
\usepackage{booktabs} 
\usepackage{siunitx} 
\usepackage{xcolor}  
\usepackage[hidelinks]{hyperref}
\hypersetup{
  pdftitle={Flat-Top Solitons as Acoustic Resonant Cavities},
  pdfauthor={M. O. D. Alotaibi, L. Al Sakkaf, and U. Al Khawaja}  
}            
\newcommand{\dd}{\mathrm{d}} 
\newcommand{\ii}{\mathrm{i}}
\newcommand{\lap}{\nabla^{2}}

\begin{document} 
\title{Flat-Top Solitons as Acoustic Resonant Cavities}
\author{M. O. D. Alotaibi}
\email{majed.alotaibi@ku.edu.kw}
\affiliation{Department of Physics, College of Science, Kuwait University,
Sabah Al Salem University City, P.O. Box 5969, Safat 13060, Shadadiya, Kuwait}

\author{L. Al Sakkaf}
\affiliation{Department of Applied Sciences, College of Engineering,
Abu Dhabi University, Al-Ain 59911, United Arab Emirates}

\author{U. Al Khawaja}
\affiliation{Department of Physics, School of Science, The University of Jordan,
Amman 11942, Jordan}
\affiliation{Department of Physics, United Arab Emirates University,
P.O. Box 15551, Al-Ain, United Arab Emirates}

\begin{abstract}
A two-dimensional flat-top soliton forms an acoustic resonator without external confinement. Its nearly uniform interior carries sound, while its moving edge selects a discrete sequence of radial wavelengths. For the cubic--quintic nonlinear Schr\"odinger equation, we connect this spectrum to bulk compressibility, radius, and line tension. Numerical simulations confirm the predicted resonances below the particle-emission threshold. The soliton itself provides the sound-carrying medium and the moving boundary, allowing its resonance frequencies to be predicted from its stationary properties.
\end{abstract}

\maketitle

In an ordinary acoustic cavity, a material medium carries sound
and surrounding walls select its resonant frequencies~\cite{Kinsler2000,Rayleigh2011,Rona2007}. In a
flat-top soliton, the medium and its moving boundary arise
together from the same nonlinear field~\cite{Michinel2002,Baizakov2011,Kartashov2019}. The nonlinear balance
that localizes the field creates a nearly uniform interior
surrounded by a narrow edge. The interior carries sound, while
the edge moves in response to pressure changes. How do the
interior and the moving edge combine to determine the resonance
frequencies when neither the size nor the edge motion is
externally imposed?
 
We answer this question for the two-dimensional
cubic--quintic nonlinear Schr\"odinger (CQ-NLS) equation.
We derive a radial quantization condition that connects
the resonance frequencies to the soliton's sound speed,
radius, and line tension, and test it using numerical
simulations. The central result is a quantitative link
between the soliton's stationary properties and its internal
resonance frequencies. The moving edge selects discrete
radial wavelengths from those allowed by the bulk dispersion
relation, turning the soliton into an acoustic cavity.

Competing attractive and repulsive nonlinearities make the CQ-NLS
equation a simple setting for this problem. Broad localized
states develop a finite-density interior and a smooth transition
to vacuum, motivating their description as liquid light
condensates~\cite{Michinel2002,Novoa2009}. Their collisions
reveal the roles of internal excitation, radiation, and
interfacial energetics~\cite{AlKhawaja2024,Alotaibi2026}.
Here, we focus on the natural frequencies at which an individual
soliton oscillates after a brief disturbance.

Small oscillations of CQ beams~\cite{Michinel2002}, breathing
modes of self-bound quantum droplets~\cite{Sturmer2021}, and
surface oscillations of compressible droplets~\cite{Mitsuhashi2026}
have been studied previously. A moving-interface theory also
describes capillary waves along the edge of two-dimensional
CQ-NLS states~\cite{Kozyreff2010}. Our focus is a sequence of
radial oscillations in which compression and expansion extend
across the soliton interior, with no angular variation.
Compression raises the interior pressure, providing the
restoring force, while line tension changes the pressure
balance at the curved edge. Together, these effects determine
the allowed radial wavelengths and their frequencies.

We test the analytical prediction using numerical simulations.
A brief potential pulse excites the soliton, and we measure its
oscillation frequencies after the pulse is removed.
The numerical simulations include the soliton's smooth edge,
allowing us to check how well the sharp-interface approximation
predicts the oscillation frequencies.
The measured frequencies agree closely with the predictions
and lie below the particle-emission threshold. Their spatial
patterns show concentric regions of compression and expansion,
supporting their interpretation as standing sound waves
inside the soliton.


\textit{Stationary state and pressure balance}---We consider
\begin{equation}
 \ii\hbar\partial_t\Psi=-\frac{\hbar^2}{2m}\lap\Psi
 +(g_2|\Psi|^2+g_3|\Psi|^4)\Psi,
 \label{eq:cqnlse}
\end{equation}
with $g_2<0$, $g_3>0$, density $n=|\Psi|^2$, and radial
Laplacian $\lap=\partial_r^2+r^{-1}\partial_r$.
Attraction binds the soliton, while repulsion resists
compression. A stationary state is
$\Psi=u(r)e^{-\ii\mu t/\hbar}$, with $u'(0)=0$,
$u(r)\to0$ as $r\to\infty$, and conserved norm
$N=2\pi\int_0^\infty u^2r\,\dd r$. A broad soliton
has nearly uniform density $n_0$ inside a radius $R$.

To find how the interior responds to compression, we
derive its pressure from the bulk interaction energy
density and local chemical potential,
\begin{equation}
 \mathcal E(n)=\frac{g_2}{2}n^2+\frac{g_3}{3}n^3,
 \qquad \mu_{\rm loc}(n)=g_2n+g_3n^2.
 \label{eq:equation_of_state}
\end{equation}
For a uniform region of area $\mathcal A$,
the total bulk interaction energy is
$E=\mathcal A\mathcal E(N/\mathcal A)$. Compression
reduces the area at fixed norm, raising the density.
The energy change defines the pressure as
$P=-(\partial E/\partial\mathcal A)_N
=n\mu_{\rm loc}-\mathcal E$~\cite{LandauLifshitz1980,Novoa2009}, hence
\begin{equation}
 P(n)=\frac{g_2}{2}n^2+\frac{2g_3}{3}n^3,
 \qquad P'(n_0)=n_0\alpha,
 \label{eq:pressure}
\end{equation}
where the prime denotes differentiation with respect to density
(evaluated here at $n=n_0$), $\alpha=g_2+2g_3n_0>0$,
and $n_0$ is the equilibrium density.
Compression therefore
raises the pressure, providing the restoring force for sound.

The edge carries a line tension $\sigma$, an energy per
unit length. In the sharp-interface approximation with
constant $\sigma$, pressure and line tension are related
at mechanical equilibrium through the
two-dimensional Young--Laplace relation
\begin{equation}
 P(n_0)=\frac{\sigma}{R}.
 \label{eq:equilibrium}
\end{equation}
This balance connects the interior density, radius, and
line tension. As $R\to\infty$, the pressure tends to zero
while the interior density approaches a finite value.
The pressure derivation, fixed-norm equilibrium argument,
and coexistence values are given in Supplemental Material,
Eqs.~\ref*{SM-eq:edens}--\ref*{SM-eq:bulk_eq}~\cite{Supplemental}.


\textit{From field dynamics to radial sound}---Writing
$\Psi=\sqrt n\,e^{\ii S}$ separates density and phase,
with fluid velocity
$\bm v=(\hbar/m)\bm\nabla S$~\cite{Dalfovo1999,PitaevskiiStringari2016}.
For a small density disturbance $n=n_0+\eta$ in a
uniform interior at rest, particle conservation and
fluid acceleration give
\begin{align}
 \partial_t\eta+n_0\bm\nabla\!\cdot\delta\bm v&=0,
 \label{eq:continuity}\\
 m\partial_t\delta\bm v&=-\bm\nabla\delta\mu.
 \label{eq:euler}
\end{align}
The chemical-potential disturbance contains an
interaction term and a term from the energy cost
of spatial density variations:
$
 \delta\mu=\alpha\eta-\beta\lap\eta, \beta=\hbar^2/(4mn_0).
$
Taking a time derivative of Eq.~\eqref{eq:continuity}
and using Eq.~\eqref{eq:euler} eliminates the velocity:
\begin{equation}
 \partial_t^2\eta=\frac{n_0}{m}\lap\delta\mu
 =\frac{n_0}{m}\lap(\alpha\eta-\beta\lap\eta).
 \label{eq:master}
\end{equation}
We seek radial compressional-wave solutions of the form
$\eta(r,t)=F(r)e^{-\ii\omega t}$, where
$F(r)$ describes the radial profile and $\omega$
is the angular frequency. Substitution gives
$-\omega^2F=(n_0/m)[\alpha\lap F-\beta\nabla^4F]$,
where $\nabla^4F=\lap(\lap F)$.
This equation factors as
$(\lap+q^2)(\lap-\kappa^2)F=0$, with
$\kappa^2-q^2=\alpha/\beta$ and
$q^2\kappa^2=m\omega^2/(n_0\beta)$.
Here $q,\kappa>0$, and $q$ is the radial wave number.
The full derivation, including both radial branches
and the approximation in which we keep only the oscillating
component, is given in Supplemental Material,
Eqs.~\ref*{SM-eq:app_separated_mode}--\ref*{SM-eq:app_bessel_mode_final}~\cite{Supplemental}.

The component $F_q$ associated with the first factor
satisfies
\begin{equation}
 F_q''+\frac{1}{r}F_q'+q^2F_q=0.
 \label{eq:bessel_mode}
\end{equation}
Here primes denote radial derivatives.
This is Bessel's equation of order zero, with
independent solutions $J_0(qr)$ and $Y_0(qr)$.
We exclude $Y_0$ because it diverges at the center,
leaving $F_q(r)=\eta_qJ_0(qr)$, where $\eta_q$
sets the amplitude.

The relations defining $q$ and $\kappa$ give
\begin{equation}
 \omega^2=\frac{n_0}{m}q^2(\alpha+\beta q^2)
 =c_s^2q^2+\frac{\hbar^2q^4}{4m^2},
 \label{eq:dispersion}
\end{equation}
where $c_s^2=n_0\alpha/m=P'(n_0)/m$.
The first term describes sound driven by pressure.
The second term provides an additional restoring effect
due to spatial density variations, increasing the frequency
at a given wave number. Defining $\xi^2=\beta/\alpha$,
this correction is small when $q\xi\ll1$, meaning that
the density varies over distances much larger than $\xi$.
In this long-wavelength limit, $\omega\simeq c_s q$,
so disturbances travel approximately at the sound speed $c_s$.
At coexistence, $\xi$ sets the width of a flat
interface, so $R\gg\xi$ describes a broad interior
with a narrow edge. The interface profile and planar line
tension are given in Supplemental Material,
Eqs.~\ref*{SM-eq:planar_front}--\ref*{SM-eq:planar_tension}~\cite{Supplemental}.

The second factor, $\lap-\kappa^2$, gives a solution
proportional to the modified Bessel function $I_0(\kappa r)$,
with $\kappa^2=q^2+\xi^{-2}$. This function remains finite
at the center and grows toward the edge without oscillating
with radial position. We keep only the spatially
oscillating $J_0$ component. Omitting $I_0$ is an
approximation that neglects a possible change near
the edge.

The bulk gives the frequency for each $q$; the moving edge selects which values are allowed, as we show in the next subsection.


\textit{The moving edge selects the wavelengths}---We now determine how the motion and pressure balance
of the soliton edge select the allowed radial wavelengths.
Our goal is to obtain the wave-number condition,
Eq.~\eqref{eq:corrected_root} below, from which we
calculate the resonance frequencies. To this end, we approximate the smooth soliton edge by a sharp circular
boundary at $r=R+\zeta(t)$, where $R$ is the equilibrium
radius and $\zeta(t)$ is a small radial displacement,
$|\zeta|\ll R$. We assume that the line tension $\sigma$
remains constant during the oscillation. To connect the
density wave inside the soliton to the motion of its edge,
we need two boundary conditions.

First, the edge moves with the fluid. Its speed,
$\dot\zeta(t)$, must therefore equal the radial fluid
velocity at the moving boundary. The equilibrium fluid
is at rest, so this velocity is the disturbance
$\delta v_r$. To first order, we evaluate this velocity at the
equilibrium radius $R$.

Second, line tension pulls the curved edge inward, while
the interior pressure pushes it outward. We assume that
these effects balance at each instant according to the
Young--Laplace relation: the pressure difference across
the edge equals line tension times curvature. The exterior
vacuum has zero pressure, and the curvature of the circular
edge is $1/(R+\zeta)$. These two conditions---the edge moves with the fluid
and the interior pressure balances line tension---give
(Supplemental Material,
Eqs.~\ref*{SM-eq:app_material_edge_exact}--\ref*{SM-eq:edge}~\cite{Supplemental}):
\begin{equation}
 \delta v_r(R,t)=\dot\zeta(t),\qquad
 P(R+\zeta,t)=\frac{\sigma}{R+\zeta}.
 \label{eq:edge}
\end{equation}

At equilibrium, $\zeta=0$ and $P_0=\sigma/R$.
For a small displacement, the pressure balance gives
$\delta P(R,t)=-\sigma\zeta(t)/R^2$ to first order
(Supplemental Material,
Eq.~\ref*{SM-eq:capillary_pressure}~\cite{Supplemental}).
Thus, an outward displacement lowers the pressure
needed to balance line tension.

A small density change $\eta=n-n_0$ changes the pressure
by $\delta P=n_0\alpha\eta$ (Supplemental Material,
Eq.~\ref*{SM-eq:app_bulk_pressure_expansion}~\cite{Supplemental}).
Both expressions must agree at the edge: the density wave
must produce the pressure change required by the edge
motion. Together with the fluid equations, this requirement
selects the allowed radial waves.

\begin{figure}[t]
 \includegraphics[width=\columnwidth]{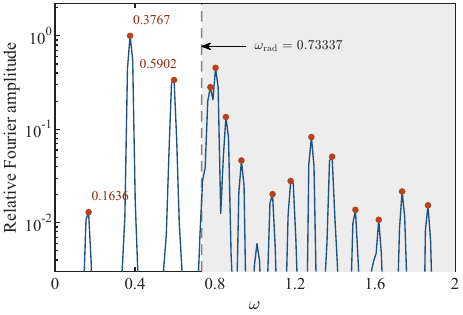}
 \caption{Fourier spectrum of the density near the soliton center
 after the pulse is removed for $N=800$. The horizontal axis shows
 the angular frequency $\omega$; the vertical axis shows the Fourier
 amplitude normalized to its maximum, on a logarithmic scale.
 Circles mark detected peaks; labels identify the three
 frequencies compared with theory. The vertical dashed line marks
 the particle-emission threshold $\omega_{\rm rad}=|\mu|/\hbar=0.73337$
 in the numerical units $\hbar=m=1$.
 In the shaded region, waves can propagate into the surrounding vacuum.}
 \label{fig:direct_spectrum}
\end{figure}

The edge moves with the fluid, so its acceleration equals
the fluid acceleration at the edge. For an oscillation
at frequency $\omega$, the edge acceleration is
$-\omega^2\zeta(t)$. The fluid equation,
Eq.~\eqref{eq:euler}, therefore relates the edge
displacement to the density wave. Substituting this
relation into the pressure balance removes $\zeta$,
leaving a condition on the density profile alone
(Supplemental Material,
Eqs.~\ref*{SM-eq:harmonic_edge}--\ref*{SM-eq:app_robin_final}~\cite{Supplemental}).
For the profile used here, $F(r)=\eta_qJ_0(qr)$,
using the bulk dispersion relation,
Eq.~\eqref{eq:dispersion}, gives
\begin{equation}
 J_0(z)-\frac{\lambda}{z}J_1(z)=0,\qquad
 z=qR,\quad\lambda=\frac{\sigma}{n_0^2\alpha R}.
 \label{eq:corrected_root}
\end{equation}
Here $\lambda$ compares the edge-pressure scale $\sigma/R$
with the bulk compression scale $n_0^2\alpha$.
Only isolated positive values of $z$
satisfy this condition, selecting the allowed radial
wavelengths
(Supplemental Material,
Eqs.~\ref*{SM-eq:app_bessel_edge_values}--\ref*{SM-eq:app_root_final}~\cite{Supplemental}).

Each positive root $z_n$ of Eq.~\eqref{eq:corrected_root}
selects a radial wave number $q_n=z_n/R$. Substituting
this into the bulk dispersion relation gives
\begin{equation}
 q_n=\frac{z_n}{R},\qquad
 (\omega_n^{\rm th})^2=c_s^2\frac{z_n^2}{R^2}
 +\frac{\hbar^2z_n^4}{4m^2R^4}.
 \label{eq:corrected_spectrum}
\end{equation}
Here $\omega_n^{\rm th}$ is the predicted angular frequency of the
$n$th radial mode.
The first term describes sound; the second is an
additional restoring effect from density gradients.
Line tension changes the allowed roots $z_n$.
Thus, the stationary soliton's density, radius, and
line tension determine its predicted resonance frequencies
(Supplemental Material,
Eqs.~\ref*{SM-eq:app_frequency_dimensional}--\ref*{SM-eq:app_scaled_spectrum}~\cite{Supplemental}).

For a broad soliton, the line-tension shift becomes small.
If also $q_n\xi\ll1$, the density-gradient contribution
is negligible, and the frequencies approach
\begin{equation}
 \omega_n^{\rm th}\simeq\frac{c_sj_{0,n}}{R},
 \label{eq:acoustic_limit}
\end{equation}
where $j_{0,n}$ is the $n$th positive zero of $J_0$.
This is the acoustic-cavity result: sound speed and
radius set the frequency scale, while the Bessel zeros
set the frequency ratios (Supplemental Material,
Eqs.~\ref*{SM-eq:app_lambda_large_radius}--\ref*{SM-eq:app_acoustic_cavity_limit}~\cite{Supplemental}).

These predictions assume a sharp edge, a uniform interior,
constant line tension, and use of only the $J_0$ density
profile. We now test them using numerical simulations.

\begin{figure*}[t]
 \centering\includegraphics[width=\textwidth]{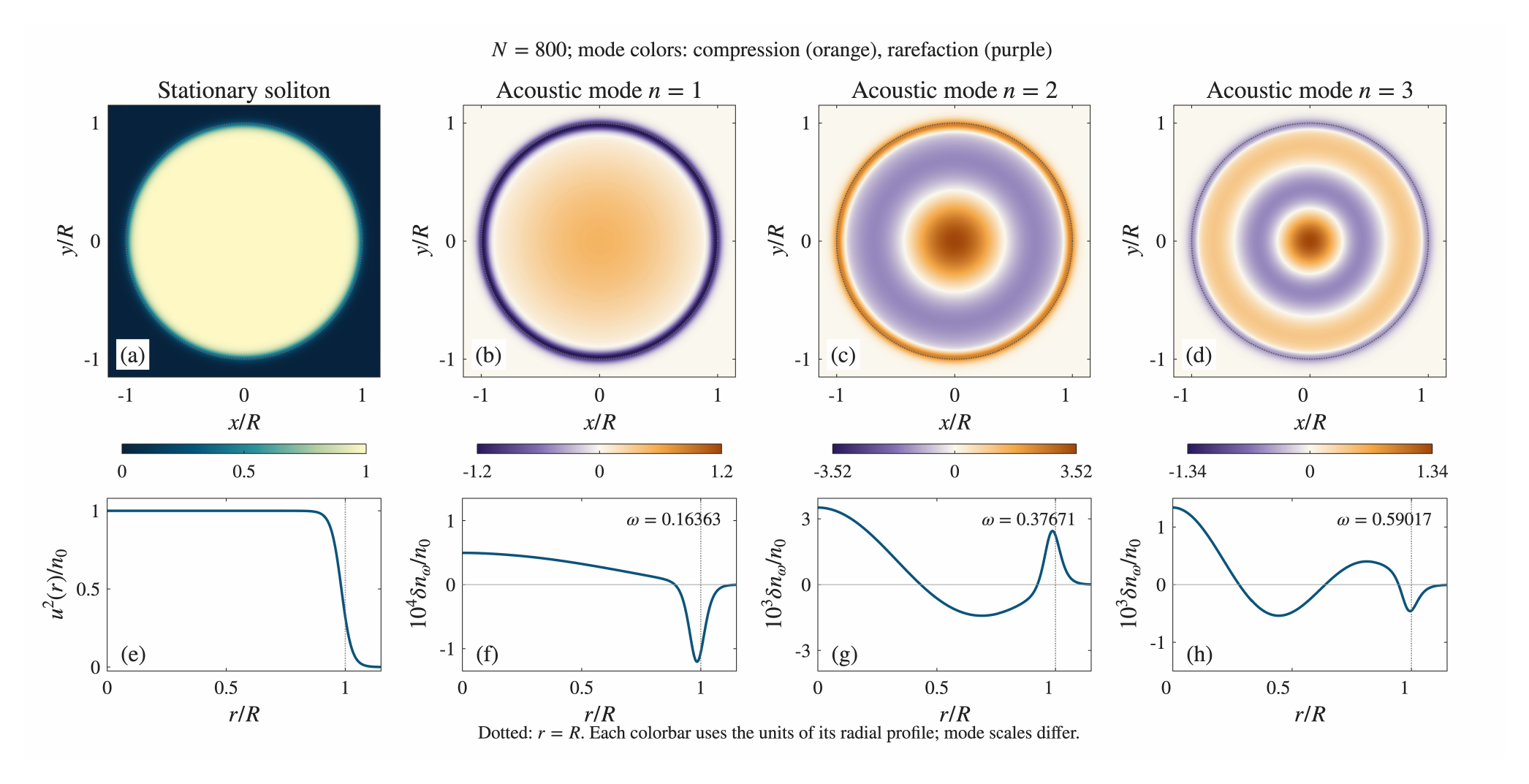}
 \caption{Equilibrium density and three radial oscillations for
 $N=800$. Each column pairs a top view (a)--(d) with its radial
 profile (e)--(h). The first column shows the stationary soliton;
 the remaining columns show the three frequency components
 extracted from the response after the pulse is removed
 (Fig.~\ref{fig:direct_spectrum}).
 For the oscillations, positive and negative values represent
 density changes relative to the stationary soliton.
 Orange indicates increased density
 (compression), purple decreased
 density (rarefaction), and the pale color zero change. Each
 component is shown at its own phase of central compression;
 the panels do not represent one common instant.
 To make the small density changes visible, we enlarge them by a
 factor of $10^4$ in panel (f) and $10^3$ in panels (g) and (h).
 Each colorbar uses the same scale as the corresponding radial
 profile; its limits are chosen separately for each component.
 Dotted circles and lines mark the soliton radius $R$.
 Three-dimensional views in Supplemental Fig.~\ref*{SM-fig:mode_gallery}
 show the same patterns as raised and lowered surfaces.}
 \label{fig:driven_density}
\end{figure*}

\textit{Numerical test of the predicted resonances}---We test
Eq.~\eqref{eq:corrected_spectrum} using numerical simulations
of Eq.~\eqref{eq:cqnlse}. For $\hbar=m=1$, $g_2=-4$, $g_3=4$,
and $N=800$, we obtain the stationary soliton using the
imaginary-time-propagation method. The equilibrium central density
$n_0=0.75818$, radius $R=18.61865$, and line tension
$\sigma=0.23448$, calculated numerically from the stationary
profile, determine the roots of Eq.~\eqref{eq:corrected_root}
and hence the predicted frequencies in
Eq.~\eqref{eq:corrected_spectrum}. Their definitions and calculation
are given in Supplemental Material, Sec.~\ref*{SM-sec:stationary} C~\cite{Supplemental}.

To excite radial oscillations, we apply a weak, brief radial
potential pulse to the stationary soliton. We use the split-step
method to simulate its response and measure the oscillation
frequencies after the pulse is removed. The pulse parameters,
numerical settings, and Fourier analysis are summarized in
Supplemental Material, Sec.~\ref*{SM-sec:numerics}~\cite{Supplemental}.

Figure~\ref{fig:direct_spectrum} shows the resulting spectrum.
The first three measured resonances agree closely with
Eq.~\eqref{eq:corrected_spectrum}: their differences from the
predictions are $+1.29\%$, $+0.87\%$, and $-0.22\%$
(Table~\ref{tab:frequencies}). These frequencies lie below the
particle-emission threshold,
$\omega_{\rm rad}=|\mu|/\hbar=0.73337$~\cite{Petrov2015,Tylutki2020}.
Its derivation is given in Supplemental Material,
Sec.~\ref*{SM-sec:high_frequency}, Eq.~\ref*{SM-eq:emission_threshold}~\cite{Supplemental}.
At these frequencies, small-amplitude waves cannot propagate into
the surrounding vacuum, supporting the picture of sound confined
within the soliton. Above this threshold, waves can propagate outside the soliton,
so deviations between the confined-cavity prediction and the
numerical results are expected.

\begin{table}[!htbp]
 \caption{Predicted [Eq.~\eqref{eq:corrected_spectrum}] and measured
 radial oscillation frequencies for $N=800$, with $n$ denoting the
 radial mode index. The percentage difference is
 $100(\omega_n^{\rm num}-\omega_n^{\rm th})/\omega_n^{\rm th}$.}
 \label{tab:frequencies}
 \begin{ruledtabular}
 \begin{tabular}{cccc}
 $n$ & $\omega_n^{\rm th}$ & $\omega_n^{\rm num}$ & Difference ($\%$)\\
 1 & 0.16155 & 0.16363 & $+1.29$\\
 2 & 0.37347 & 0.37671 & $+0.87$\\
 3 & 0.59149 & 0.59017 & $-0.22$\\
 \end{tabular}
 \end{ruledtabular}
\end{table}

Figure~\ref{fig:driven_density} shows the stationary soliton and
the density patterns at the three measured resonance frequencies.
Panels (a)--(d) show top views, with the corresponding radial
profiles in panels (e)--(h). Panels (a) and (e) show the stationary
density, while the pairs (b,f), (c,g), and (d,h) show the density
changes for the first, second, and third resonances, respectively.
Panel (e) shows the stationary density normalized by $n_0$;
panels (f)--(h) show the density changes relative to the stationary
profile, also normalized by $n_0$. To make the small density changes
visible, we enlarge them by a factor of $10^4$ in panel (f) and
$10^3$ in panels (g) and (h). Orange and purple regions indicate
compression and rarefaction; these exchange places after half an
oscillation period. The interior patterns support their
interpretation as radial standing sound waves. The corresponding
three-dimensional views are shown in Supplemental Material,
Fig.~\ref*{SM-fig:mode_gallery}~\cite{Supplemental}.


\textit{Conclusion}---A flat-top soliton acts as an acoustic cavity
whose confinement is supplied by its own moving edge. Within a
sharp-interface description, pressure balance at this edge selects
the allowed radial wavelengths and connects the stationary soliton's
properties to its internal resonance frequencies. The bulk
compressibility determines the sound speed, the radius sets the
wavelength scale, and line tension modifies the boundary condition.
The frequencies measured in numerical simulations after the pulse
is removed agree closely with the analytical prediction below the
particle-emission threshold. Their spatial patterns support an
interpretation as radial standing sound waves. These results connect
the internal dynamics of a self-bound soliton to its stationary
structure and provide a basis for studying how compressional
resonances change with soliton size and interface thickness.

\bibliography{references}
\end{document}